\pdfoutput=1

\documentclass[11pt]{article}

\usepackage{acl}

\usepackage{times}
\usepackage{latexsym}

\usepackage[T1]{fontenc}

\usepackage[utf8]{inputenc}

\usepackage{microtype}

\usepackage{amsmath,amsfonts,bm}

\def\eqref#1{equation~\ref{#1}}

\def\1{\bm{1}}

\DeclareMathAlphabet{\mathsfit}{\encodingdefault}{\sfdefault}{m}{sl}
\SetMathAlphabet{\mathsfit}{bold}{\encodingdefault}{\sfdefault}{bx}{n}

\usepackage{xspace}
\usepackage{hyperref}
\usepackage{url}
\usepackage{wrapfig}
\usepackage{graphicx}
\usepackage{caption}
\usepackage{subcaption}
\usepackage{multirow}
\usepackage{booktabs}
\usepackage{flushend}

\newcommand{\eg}{\hbox{\emph{e.g.,}}\xspace}
\newcommand{\ie}{\hbox{\emph{i.e.,}}\xspace}
\newcommand{\wrt}{\hbox{\emph{w.r.t.}}\xspace}

\newcommand{\tool}{{\sc DISCO}\xspace}
\newcommand{\toolM}{{\sc DISCO}$_{medium}$\xspace}
\newcommand{\toolS}{{\sc DISCO}$_{small}$\xspace}

\newcommand{\mlm}{MLM}
\newcommand{\mlmclr}{MLM+CLR$^+$}
\newcommand{\mlmclrpn}{MLM+CLR$^\pm$\xspace}
\newcommand{\tmlmclrpn}{MLM+CLR$^\pm$+NT-MLM\xspace}

\title{Towards Learning (Dis)-Similarity of Source Code from Program Contrasts}

\author{Yangruibo Ding$^\S$, Luca Buratti$^\dagger$, Saurabh Pujar$^\dagger$, Alessandro Morari$^\dagger$, \\
\textbf{Baishakhi Ray$^\S$}, \textbf{and Saikat Chakraborty$^\S$} \\
$^\S$ Columbia University \\
$^\dagger$ IBM Research \\
$^\S$\texttt{\{yrbding, rayb, saikatc\}@cs.columbia.edu} \\
$^\dagger$\texttt{\{luca.buratti1, saurabh.pujar\}@ibm.com, amorari@us.ibm.com}
  }

\begin{document}
\maketitle
\begin{abstract}

Understanding the functional (dis)-similarity of source code is significant for code modeling tasks such as software vulnerability and code clone detection.
We present \tool(\emph{DIS-similarity of COde}), a novel self-supervised model focusing on identifying (dis)similar functionalities of source code. Different from existing works, our approach does not require a huge amount of randomly collected datasets. Rather, we design structure-guided code transformation algorithms to generate synthetic code clones and inject real-world security bugs, augmenting the collected datasets in a targeted way. We propose to pre-train the Transformer model with such automatically generated program contrasts to better identify similar code in the wild and differentiate vulnerable programs from benign ones. To better capture the structural features of source code, we propose a new cloze objective to encode the local tree-based context (\eg parents or sibling nodes). We pre-train our model with a much smaller dataset, the size of which is only 5\% of the state-of-the-art models' training datasets, to illustrate the effectiveness of our data augmentation and the pre-training approach. The evaluation shows that, even with much less data, \tool can still outperform the state-of-the-art models in vulnerability and code clone detection tasks.

\end{abstract}

\section{Introduction}
\label{sec1:intro}

Understanding the functional similarity/dissimilarity of source code is at the core of several code modeling tasks such as software vulnerability and code clone detection, which are important for software maintenance~\cite{kim2017vuddy, li2016vulpecker}.
Existing pre-trained Transformer models~\citep{guo2021graphcodebert, feng2020codebert, ahmad2021plbart} show promises for understanding code syntax (\ie tokens and structures). However, they still get confused when trying to identify functional (dis)-similarities. For instance, syntax-based models can embed two code fragments with identical functionality but very different tokens and structures as distinct vectors and fail to identify them as semantically similar.  
Likewise, these models cannot distinguish between two code fragments that differ in functionalities but share a close syntactic resemblance. For example, consider an if statement {\tt if(len(buf) $<$ N)} checking buffer length before accessing the buffer. Keeping the rest of the program the same, if we simply replace the token `$<$' with `$\leq$,' the modification can potentially trigger security vulnerability, \eg buffer overflow bug\footnote{https://en.wikipedia.org/wiki/Buffer\_overflow}. It is challenging for existing pre-training techniques to tell apart such subtle differences in the functionalities.

In addition, existing pre-training techniques rely on a huge volume of training corpus that is randomly selected. For fine-tuning tasks like code clone detection or vulnerability detection, such random selection of training data is never tailored to teach the model about code functionalities.

To address these limitations, we present \tool, a self-supervised pre-trained model that jointly learns the general representations of source code and specific functional features for identifying source code similarity/dis-similarity.
Similar to state-of-the-art pre-trained Transformer models~\citep{devlin-etal-2019-bert, liu2019roberta}, we apply the standard masked language model (MLM) to capture the token features of source code. 
To learn about the structural code properties, we propose a new auxiliary pre-training task that consumes additional inputs of local tree-based contexts (\eg parent or sibling nodes in abstract syntax trees) and embeds such structural context, together with the token-based contexts, into each token representation. On top of such well-learned general code representations, we further incorporate prior knowledge of code clones and vulnerable programs into the pre-training to help the model learn the functional (dis)-similarity. We design structure-guided code transformation heuristics to automatically augment each training sample with one synthetic code clone (\ie positive samples) that is structurally different yet functionally identical and one vulnerable contrast (\ie hard negative samples) that is syntactically similar but injected with security bugs. 
During the pre-training, \tool learns to bring similar programs closer in the vector space and differentiate the benign code from its vulnerable contrast, using a contrastive learning objective. Since we augment the dataset in a more targeted way than existing works and the model {\em explicitly} learns to reason about a code \wrt its functional equivalent and different counterparts during pre-training, \tool can learn sufficient knowledge for downstream applications from a limited amount of data, consequently saving computing resources. In particular, we evaluate \tool for clone detection and vulnerability detection, as the knowledge of similar/dissimilar code fragments is at the core of these tasks.  

\vspace{-0.5mm}
To this end, we pre-train \tool on a \emph{small} dataset, with only 865 MB of C code and 992 MB Java code from 100 most popular GitHub repositories, and evaluate the model on four different datasets for vulnerability and code clone detection. Experiments show that our small models outperform baselines that are pre-trained on 20$\times$ larger datasets. 
The ablation study (\S \ref{subsec:medium_model}) also reveals that pre-training our model with 10$\times$ larger datasets further improves the performance up to 8.2\%, outperforming state-of-the-art models by 1\% for identifying code clones and up to 9.6\% for vulnerability detection, even if our dataset is still smaller.

\vspace{-0.5mm}
In summary, our contributions are: 1) We design structure-guided code transformation heuristics to automatically augment training data to integrate prior knowledge of vulnerability and clone detection without human labels. 2) We propose a new pre-training task to embed structural context to each token embedding. 3) We develop \tool, a self-supervised pre-training technique that jointly and efficiently learns the textual, structural, and functional properties of code. Even though pre-trained with significantly less data, \tool matches or outperforms the state-of-the-art models on code clone and vulnerability detection.
\vspace{-1mm}
\section{Related Works}
\label{sec2:related}
\textbf{Pre-training for Source Code.}
Researchers have been passionate about pre-training Transformer models~\citep{vaswani2017transformer} for source code with two categories: encoder-only and encoder-decoder~\citep{ahmad2021plbart, wang2021codet5,roziere2021dobf,phan2021cotext}. Our work focuses on pre-training encoder-only Transformer models to understand code. Existing models are pre-trained with different token level objectives, such as
masked language model (MLM) \citep{cubert, buratti2020cbert}, next sentence prediction (NSP) \citep{cubert}, replaced token detection, and bi-modal learning between source code and natural languages \citep{feng2020codebert}. However, these approaches ignore the underlying structural information to fully understand the syntax and semantics of programming languages. Recently, more works aimed to understand the strict-defined structure of source code leveraging abstract syntax tree (AST) \citep{zuegner_code_transformer_2021, Jiang2021TreeBERT}, control/data flow graphs (CFG/DFG) \citep{guo2021graphcodebert}. \tool leverages code structures differently from existing works in two ways: (a.) with AST/CFG/DFG, we automatically generate program contrasts to augment the datasets targeting specific downstream tasks. (b.) \tool takes an additional input of local AST context, and we propose a new cloze task to embed local structural information into each token representation.

\vspace{-1mm}
\noindent\textbf{Self-supervised Contrastive Learning.} Self-supervised contrastive learning, originally proposed for computer vision~\citep{chen20simclr}, has gained much interest in language processing~\citep{giorgi-etal-2021-declutr, wu2020clear, gao2021simcse}. 
The common practice of self-supervised contrastive learning is building similar counterparts, without human interference, for the original samples and forcing the model to recognize such similarity from a batch of randomly selected samples. 
Corder~\citep{bui2021corder} leverages contrastive learning to understand the similarity between a program and its functionally equivalent code. While Corder approach will help code similarity detection type of applications, their pre-training does not learn to differentiate syntactically very close, but functionally different programs. 
Such differentiation is crucial for models to work well for bug detection~\citep{ding2020patching}. ContraCode~\citep{jain2020contrastive} also leverages contrastive learning. However, they generate negative contrast for a program from unrelated code examples, not from variants of the same code. They also do not encode the structural information into the code as we do. Inspired by the empirical findings that hard negative image and text samples are beneficial for contrastive learning~\citep{gao2021simcse, robinson2021hardnegative}, \tool learns both from equivalent code as the positive contrast, and functionally different yet syntactically close code as the \emph{hard-negative} contrast. 
We generate hard-negative samples by injecting small but crucial bugs in the original code (\S \ref{subsec:bug_injection}). 

\section{Data Augmentation Without Human Labels}
\label{sec3:data_aug}
\begin{figure*}
    \centering
    \begin{subfigure}[b]{0.32\textwidth}
        \centering
        \includegraphics[width=\textwidth]{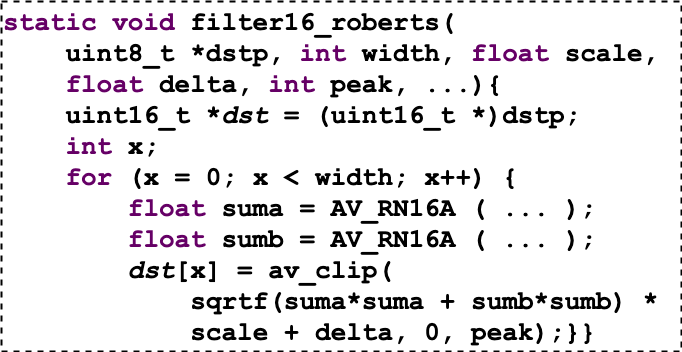}
        \caption{Original Code}
        \label{fig:orig_sample}
    \end{subfigure}
    \begin{subfigure}[b]{0.32\textwidth}
        \centering
        \includegraphics[width=\textwidth]{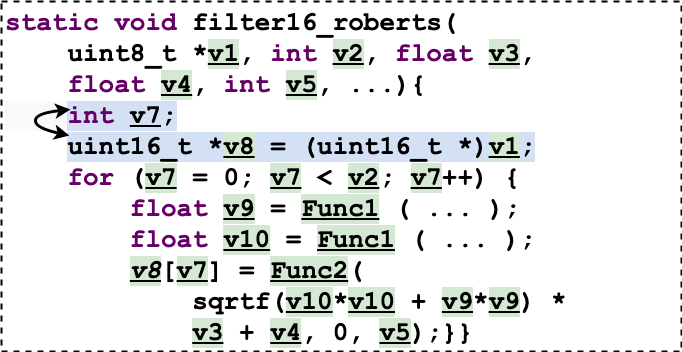}
        \caption{Functionally Equivalent Code}
        \label{fig:pos_sample}
    \end{subfigure}
    \begin{subfigure}[b]{0.32\textwidth}
        \centering
        \includegraphics[width=\textwidth]{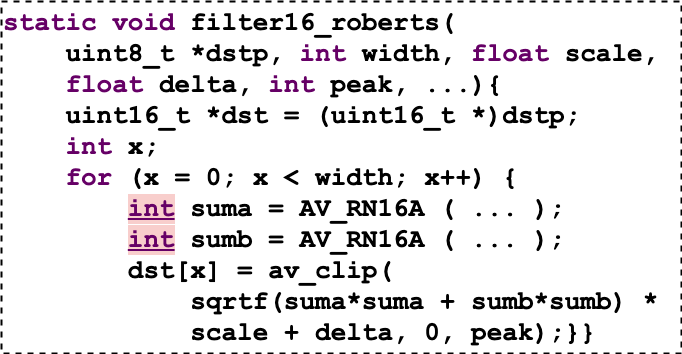}
        \caption{Bug Injected Code}
        \label{fig:neg_sample}
    \end{subfigure}
    \caption{An example illustrating data augmentation.~\ref{fig:orig_sample} shows the original code that is adapted from the CVE-2021-38094 patch. 
\ref{fig:pos_sample} shows functionality equivalent code of ~\ref{fig:orig_sample} where the original code is transformed by renaming and statements permutation. \ref{fig:neg_sample} shows a variation from~\ref{fig:orig_sample} where a potential integer overflow bug is injected.}
\label{fig:data_aug}
\vspace{-5mm}
\end{figure*}

Our pre-training aims to identify similar programs that can be structurally different (positive sample) and differentiate the buggy programs (negative sample) that share structural resemblances with the benign ones. Thus, we need a labeled positive and a negative example for each original sample. Manually collecting them 
is expensive, especially at the scale of  pre-training. To this end, we design code transformation heuristics to automatically generate such positive and negative samples so that the transformation can be applied to any amount of programs without human efforts. 

We first represent a code sample as Abstract Syntax Tree (AST), and build a control/data flow graph from the AST. 
The code transformation heuristics are then applied to this graph. 
For every original code sample ($x$), we apply semantic preserving transformation heuristics (\S\ref{subsec:similar_code_generation}) to generate a positive sample ($x^+$) and a bug injection heuristics (\S \ref{subsec:bug_injection}) to generate a hard-negative code example ($x^-$). 
We design the heuristics in a way that makes $x^+$ be the functional equivalent or semantic clone of $x$  and $x^-$ be the buggy/noisy version of $x$. 
Noted that not all heuristics are applicable to all code samples; we decide on applicable heuristics based on the flow graph of the original code. 
Figure~\ref{fig:data_aug} shows an example of the code transformation. 

\subsection{Bug Injection}
\label{subsec:bug_injection}

To generate a hard negative sample ($x^-$) from a given code ($x$), we define six categories of bug injection heuristics.
Here our goal is to maintain maximum token-level similarity to the original code, so that the model can learn to analyze source code beyond token-level similarity. 
These heuristics are inspired by the buggy code patterns from a wide range of Common Weakness Enumeration (CWE) types (Appendix~\ref{appendix:bug_injection_cwe}). 
While it is challenging to guarantee that $x^-$ will exhibit vulnerability or security bug, our heuristics will force $x^-$ to exhibit different functionality than $x$. Compared with a concurrent work from \citet{allamanis2021self}, our methods are significantly different. First, we focus on concrete types of security bugs that have been identified by the security experts, while they mainly target regular bugs. Second, our scope is not only bug detection but clone detection as well, and we apply contrastive learning to differentiate the code functionalities of code clones and vulnerabilities.

\noindent\textbf{Misuse of Data Type.}
Usage of the wrong data type can trigger several security flaws.
For instance, using a smaller data type (\eg~{\tt short}) to replace a larger one (\eg~{\tt long}) may result in an overflow bug (\eg \citet{cve-2021-38094}). Such errors are complicated to track since they are usually exhibited in input extremities (\ie very large or very small values). For languages allowing implicit typecasting, such an incorrect type may even cause imprecision, resulting in the unpredictable behavior of the code. We intentionally change the data types in $x$ to inject potential bugs, while ensuring the code can still be compiled (\eg we will not replace {\tt int} with {\tt char}). 

\noindent\textbf{Misuse of Pointer.}
Incorrect pointer usage is a major security concern.
Accessing uninitialized pointers may lead to unpredictable behavior.
A {\tt NULL} pointer or freed pointer could lead to Null Pointer Dereferencing vulnerability (\eg \citet{cve-2021-3449}). 
To inject such bugs, we randomly remove the initialization expression during pointer declaration, or set some pointers to {\tt NULL}. 

\noindent\textbf{Change of Conditional Statements.}
Programmers usually check necessary preconditions using {\tt if-statement} before doing any safety-critical operation. 
For instance, before accessing an array with an index, a programmer may add a condition checking the validity of the index. 
Lack of such checks can lead to buffer-overflow bugs in code (\eg \citet{cve-2020-24020}). 
We introduce bugs in the code by removing such small {\tt if-statement}s. 
In addition, we also inject bugs by modifying randomly selected arithmetic conditions--- replace the comparison operator ($<$, $>$, $\leq$, $\geq$, $==$, $!=$) with another operator, to inject potential out-of-bound access, forcing the program to deviate from its original behavior.

\noindent\textbf{Misuse of Variables.} 
When there are multiple variables present in a code scope, incorrect use of variables may lead to erroneous behavior of the program. 
Such errors are known as {\sc VarMisuse} bug~\citep{allamanis2018learning}. 
We induce code with such bugs by replacing one variable with another. 
To keep the resultant code compilable, we perform scope analysis on the AST and replace a variable with another variable reachable in the same scope. 

\noindent\textbf{Misuse of Values.}
Uninitialized variables or variables with wrong values may alter the program behaviors and consequently cause security flaws (\eg ~\citet{cve-2019-12730}). We modify the original code by removing the initializer expression of some variables. In addition, to induce the code with {\tt divide-by-zero} vulnerability, we identify the potential divisor variables from the flow graph and forcefully assign zero values to them immediately before the division. 

\noindent\textbf{Change of Function Calls.} We induce bugs in the code by randomly changing arguments of function calls. For a randomly selected function call, we add, remove, swap, or assign {\tt NULL} value to arguments, forcing the code to behave unexpectedly. 

\subsection{Similar Code Generation}
\label{subsec:similar_code_generation}

To generate positive samples ($x^+$) from a given code, we use three different heuristics. 
In this case, our goal is to generate functionally equivalent code while inducing maximum textual difference. 
These heuristics are inspired by code clone literature~\citep{funaro2010hybrid, sheneamer2018detection}.

\noindent\textbf{Variable Renaming.}
Variable renaming is a typical code cloning strategy and frequently happens during software development~\citep{ain2019syscodeclone}. 
To generate such a variant of the original code, we either (a.) rename a variable in the code with a random identifier name or (b.) with an abstract name such as \texttt{VAR\_i}~\citep{roziere2021dobf}. 
While choosing random identifier names, we only select available identifiers in the dataset. 
We ensure that both the definition of the variable and subsequent usage(s) are renamed for any variable renaming. We also ensure that a name is not used to rename more than one variable. 

\noindent\textbf{Function Renaming.}
We rename function calls with abstract names like \texttt{FUNC\_i}. By doing this, we make more tokens different compared with the original code but keep the same syntax and semantics. We do not rename 
library calls for the code (\eg \texttt{memcpy()} in C).

Noted that even if tokens like \texttt{VAR\_i} and \texttt{FUNC\_i} are rare in normal code, the model will not bias towards identifying samples with these tokens as positive samples. The reason is that, as shown in Figure~\ref{fig:model_arch}, $x^+, y^+$ and $z^+$ all potentially have these abstract tokens, but the model learns to move $EMB_{x}$ closer to $EMB_{x^+}$ and further from $EMB_{y^+}$ and $EMB_{z^+}$, regardless of the existence of abstract tokens.  

\noindent\textbf{Statement Permutation.}
The relative order among the program statements that are independent of each other can be changed without altering the code functionality.
More specifically, we focus on the variable declaration or initialization statements. 
We first conduct the dependency analysis to identify a set of local variables that do not depend on other values for initialization. Then we move their declaration statements to the beginning of the function and permute them.

\section{\tool}
\label{sec4:model}
\begin{figure*}
    \centering
    \includegraphics[width=\textwidth]{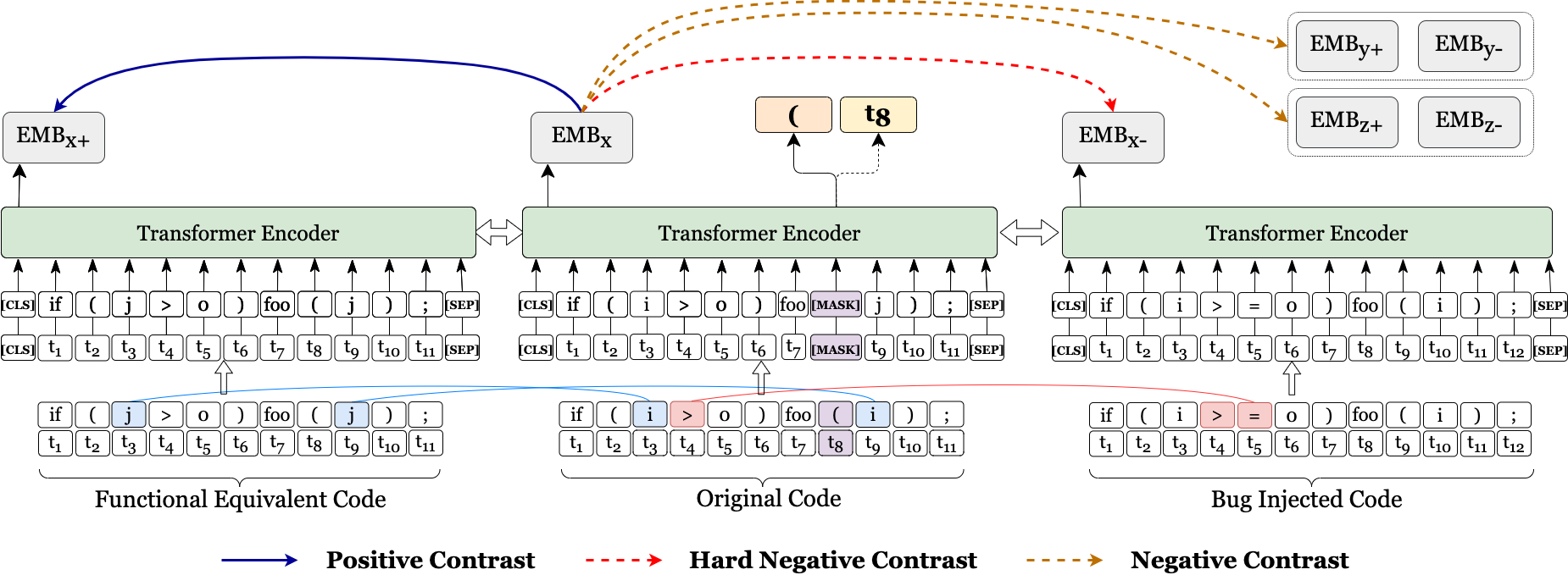}
    \caption{An illustration of \tool pre-training with a minibatch of three. The original code and its node types will be randomly masked with $[MASK]$, and the final representation of masked tokens will be used to recover their source tokens and node types. The original code, say $x$, will also be transformed to build ($x, x^+, x^-$). Then the triplet will be fed into the \textbf{same} Transformer encoder and get the embedding of each sequence with $[CLS]$ tokens for contrastive learning.}
    \label{fig:model_arch}
    \vspace{-5mm}
\end{figure*}

This section presents the model architecture, input representation, and pre-training tasks.
\tool uses a 12-layered Transformer encoder model similar to BERT.
We feed the model with both source code text and structure (AST) information (\S\ref{subsec:input_repr}). 
We pre-train \tool using three different pre-training tasks (\S\ref{subsec:pretraining-task}). 
Figure \ref{fig:model_arch} depicts an example workflow of \tool. 
We randomly select tokens in the original sample, mask them and their node types, and then use the embedding of these masks to predict them back. We further extract the sequence embeddings within a minibatch and contrast them based on the code functionality. 

\subsection{Input Representation}
\label{subsec:input_repr}
\textbf{Source Code.} Given a program ($x$), we apply a lexical analyzer to tokenize it based on the language grammar and flatten the program as a token sequence ($x_1 x_2 ... x_m$, where $x_i$ is i$^{th}$ token in the code).
We further train a sentencepiece~\citep{kudo-richardson-2018-sentencepiece} tokenizer based on such flattened code token sequences with vocabulary size 20,000. 
We use this tokenizer to divide the source code tokens into subtokens. 
We prepend the subtoken sequence with a special token {\tt [CLS]} and append with a special token {\tt [SEP]}. 
Finally, \tool converts the pre-processed code sequence $C = \{[CLS], c_1, c_2, ..., c_k, [SEP]\}$ to vectors $V^{src} = \{v_{[CLS]}^{src}, v_1^{src}, v_2^{src}, ..., v_k^{src}, v_{[SEP]}^{src}\}$ with a token embedding layer.

\noindent\textbf{Local AST Types.} 
For every token in the input code, we extract the node type ($tt$) from the syntax tree. 
Since such types are all terminal node types (\eg keyword, identifier, punctuation), we do not get enough information about the structure only with these types. 
In order to add more information about the tree, we also extract its parent type ($pt$) for each token. 
Such parent type provides us with information about the structural context of a token. For instance, when the parent type of an {\tt identifier} is {\tt Function-Declarator}, we know that the identifier is a function name. 
In contrast, when the {\tt identifier}'s parent is a {\tt Binary Expression}, it should be a variable.
Consequently, we annotate each code sub-token $c_i$ with a local AST-type token $t = tt\#pt$. 
It is worth noting that sub-tokens coming from the same code token will all have the same type. 
Therefore, we have the AST-type sequence for the code $T = \{[CLS], t_1, t_2, ..., t_k, [SEP]\}$, and \tool converts it as vectors $V^{type} = \{v_{[CLS]}^{type}, v_1^{type}, v_2^{type}, ..., v_k^{type}, v_{[SEP]}^{type}\}$ with a type embedding layer.
Appendix Table~\ref{tab: node_type} shows an example of code tokens and their AST types.
\tool generates token representation $v_i$ of subtoken $c_i$ as a sum of token embedding $v_i^{src}$ and type embedding $v_i^{type}$.
Thus, $V=V^{src} + V^{type}$.

\subsection{Pre-training}
\label{subsec:pretraining-task}
We aim to pre-train the \tool to learn the representation of source code based on (a.) token-based context, (b.) AST-based context, and (c.) code functionality. In that spirit, we pre-train \tool to optimize on three different objectives, \ie masked language model (MLM), local AST node type-MLM (NT-MLM), and Contrastive Learning (CLR).

For a given program $x$, we first embed the tokens and node-types to vectors $V = \{v_{[CLS]}, v_1, ..., v_{[SEP]}\}$. We optimize MLM loss ($\mathcal{L}_{MLM}$) (\S\ref{subsec:mlm}) and NT-MLM loss ($\mathcal{L}_{NT-MLM}$) (\S\ref{subsec:ntmlm}) based on $x$. These two loss functions learn about the textual and syntactic context of source code. For every code sample $x$ in a minibatch of input, we generate a positive example $x^+$ and a hard-negative example $x^-$ using the heuristics described in Section \ref{sec3:data_aug}. We optimize CLR loss ( $\mathcal{L}_{CLR}$) (\S\ref{subsec:clr}) considering the original code and its positive and hard-negative counterparts. 
The final loss function to optimize for pre-training \tool is 
\begin{equation*}
    \mathcal{L}(\theta) = \mathcal{L}_{MLM}(\theta) + \mathcal{L}_{NT-MLM}(\theta) + \mathcal{L}_{CLR}(\theta)
\end{equation*}


\subsubsection{Encoding Token-based Context}
\label{subsec:mlm}
We apply the standard masked language model to the original code ($x$). Given a source code sequence $C$, we randomly choose 15\% of tokens and replace them with a special token $[MASK]$ for 80\% of the time and a random token for 10\% of the time and leave the rest 10\% unchanged. We record the indices of masked token as $loc_m$, replaced token as $loc_r$ and unchanged tokens as $loc_u$ for node-type MLM. We define the union of these indices as $M = loc_m \cup loc_r \cup loc_u$. MLM will learn to recover the masked source code $\{c_i | i \in M\}$ given the Transformer encoder's output $h_i$. We present the loss for MLM as 
$\mathcal{L}_{MLM} = \sum_{i \in M} - log P(c_i | h_i)$
\subsubsection{Encoding AST-based Context } 
\label{subsec:ntmlm}

Token-based MLM re-builds the token using its surrounding tokens and successfully encodes the contextual information into each token representation. Motivated by MLM, we propose the tree-based context-aware pre-training task, to encode the structural context, such as parent, sibling, and children nodes. As we have shown in Figure~\ref{fig:model_arch}, we flatten the ASTs as sequences and we expect the flattened trees can preserve the local structure information (i.e., sub-trees containing terminal nodes), and existing work~\citep{Chakraborty2020codit, Hellendoorn2020Global} has empirically shown such potentials. To this end, we introduce AST node-type masked language model (NT-MLM). Given the corresponding AST-type sequence $T$ of source code $C$, we mask the AST types $\{t_p | p \in loc_m\}$ with the special token $[MASK]$, and replace the AST types $\{t_q | q \in loc_r\}$ with random tokens. Specifically, by doing this, we make sure that if a source code token is chosen to be masked or replaced, its corresponding AST type will perform the same operation. NT-MLM will learn to recover the masked AST type $\{t_i | i \in M\}$ given the Transformer encoder's output $h_i$. We present the loss for NT-MLM as 
$\mathcal{L}_{NT-MLM} = \sum_{i \in M} - log P(t_i | h_i)$

A recent work, CodeT5~\citep{wang2021codet5}, proposes to predict token type as well. However, our new objective is different from them in both high-level designs and the detailed implementation. First, their objective only predicts one single token type: identifiers, while our approach predicts all possible AST types. Also, we do not only consider the AST node type of tokens, but also include their AST parents to embed the local sub-tree context (\S \ref{subsec:input_repr}). Second, CodeT5 implements the identifier tagging task as a binary classification (0/1) for each token, while our NT-MLM reconstructs the local ASTs out of hundreds of distinct types.

\subsubsection{Contrastive Learning } 
\label{subsec:clr}
We adopt contrastive learning to focus on the functional characteristics of code. With the structure-guided code transformation algorithms in Section~\ref{sec3:data_aug}, we are able to generate a positive sample ($x^+$ in Figure~\ref{fig:model_arch}) and a hard negative sample ($x^-$ in Figure~\ref{fig:model_arch}) for each program in the dataset. 
More specifically, we have a minibatch of $N$ programs, and for each program, we extract the sequence representation from the Transformer outputs $\mathbf{h} = h_{[CLS]}$. We will augment every sequence in the minibatch with positive and negative samples, and then the minibatch is extended to N triplets of $(\mathbf{h}, \mathbf{h}^+, \mathbf{h}^-)$. We refer to the contrastive loss with hard negative samples from \citet{gao2021simcse} and we adapt it to our scope as follows. We use cosine similarity as the $sim()$ function and $\tau$ is the temperature parameter to scale the loss, and we use $\tau = 0.05$. 

\footnotesize
\begin{equation*}
    \mathcal{L}_{CLR} = -\log\frac{\mathbf{e}^{\text{sim}\left(\mathbf{h}, \mathbf{h}^+\right)/\tau}}{\sum^{N}_{n=1}\left(\mathbf{e}^{\text{sim}\left(\mathbf{h}, \mathbf{h}_{n}^+\right)/\tau} + \mathbf{e}^{\text{sim}\left(\mathbf{h}, \mathbf{h}_{n}^-\right)/\tau} \right)}
\end{equation*}
\normalsize
We also consider to pre-train the model with only positive counterparts as a variation. In such a case, the minibatch will contain N pairs of $(\mathbf{h}, \mathbf{h}^+)$ and the loss is computed as

\footnotesize
\begin{equation*}
    \mathcal{L}_{CLR} = -\log\frac{\mathbf{e}^{\text{sim}\left(\mathbf{h}, \mathbf{h}^+\right)/\tau}}{\sum^{N}_{n=1}\left(\mathbf{e}^{\text{sim}\left(\mathbf{h}, \mathbf{h}_{n}^+\right)/\tau} \right)}
\end{equation*}
\normalsize

\section{Experiments}
\label{sec5:expr}

In this section, we will explain our experimental settings and report the results. We evaluate our model on vulnerability and code clone detection. 

\subsection{Experimental Settings}
\label{subsec:expr_pretrain}
\textbf{Data.}
We collect our pre-training corpus from open-source C and Java projects. 
We rank Github repositories by the number of stars and focus on the most popular ones. 
After filtering out forks from existing repositories,
we collect the dataset for each language from top-100 repositories.
We only consider the ``.java'' and ``.c'' files for Java and C repositories respectively, and we further remove comments and empty lines from these files. 
The corresponding datasets for Java and C are of size of 992MB and 865MB, respectively.
Our datasets are significantly \emph{smaller} than existing pre-training models~\citep{feng2020codebert, ahmad2021plbart, guo2021graphcodebert}. 
For example, while CodeBERT and GraphCodeBERT are trained on 20GB data, we used an order of magnitude less data.  
Details of our datasets and the comparison can be found in Appendix Table~\ref{tab: pretrain_data_stats}. 

\noindent\textbf{Models.} 
To study the different design choices, we train four variations of \tool. (i) \textbf{
MLM+CLR$^{\pm}$+NT-MLM
} is trained by all three tasks with hard negative samples. (ii) \textbf{ MLM+CLR$^{\pm}$}. The input of this model only considers the source code sequence and ignores the AST-type sequence. This model helps us understand the impact of NT-MLM. 
(iii) \textbf{MLM+CLR$^+$}. This variant evaluates the effectiveness of hard negative code samples, by contrasting its performance with MLM+CLR$^{\pm}$.
(iv) \textbf{MLM}. This is the baseline trained with only MLM objective.
We provide detailed model configuration in Appendix~\ref{appendix: model_config} to ensure the reproducibility.

\noindent\textbf{Baselines.} We consider two types of baselines: encoder-only pre-trained Transformers and existing deep-learning tools designed for code clone and vulnerability detection. We do not consider encoder-decoder pre-trained Transformers as baselines, since such generative models always need much more pre-training data and training steps to converge, so it is unfair to compare our model with them. For example, PLBART uses 576G source code for pre-training, while we only use less than 1G. 
Based on the data size. As future work, we plan to pre-train the model on much larger datasets. 

\subsection{Vulnerability Detection (VD)}
\label{subsec: vul_detect}
VD is the task to identify 
security bugs: given a source code function, the model predicts 0 (benign) or 1 (vulnerable) as binary classification.

\noindent\textbf{Dataset and Metrics.} We consider two datasets for VD task: REVEAL~\citep{chakraborty2021reveal} and CodeXGLUE~\citep{msr2021codexglue, Zhou2019DevignEV}. In the real-world scenario, vulnerable programs are always rare compared to the normal ones, and ~\citet{chakraborty2021reveal} have shown such imbalanced ratio brings challenges for deep-learning models to pinpoint the bugs. To imitate the real-world scenario, they collect REVEAL dataset from Chromium (open-source project of Chrome) and Linux Debian Kernel, which keeps the ratio of vulnerable to benign programs to be roughly 1:10. Following ~\citet{chakraborty2021reveal}, we consider precision, recall and F1 as the metrics.

CodeXGLUE presents another dataset of security vulnerabilities. It is less real-world than REVEAL, since it a balanced dataset, but it has been frequently used by existing Transformer-based models to evaluate their tools for VD 
task. To compare with these baselines, we use CodeXGLUE train/valid/test splits for training and testing. We use accuracy as the metric, following the design of the benchmark.

\noindent\textbf{REVEAL.} Table~\ref{tab: reveal_result} shows the results. We compare with four deep-learning-based VD 
tools. VulDeePecker~\citep{li2018vuldeepecker} and SySeVR~\citep{li2018sysevr} apply program slices and sequence-based RNN/CNN to learn the vulnerable patterns. Devign~\citep{Zhou2019DevignEV} uses graph-based neural networks (GNN) to learn the data dependencies of program. 
REVEAL~\citep{chakraborty2021reveal} applies GNN + SMOTE~\citep{chawla2002smote} + triplet loss during training to handle the imbalanced distribution. 
We also consider pre-trained RoBERTa, CodeBERT and GraphCodeBERT, and a 12-Layer Transformer model trained from scratch. 


\begin{table}[!htb]
\centering
\caption{\small Vulnerability Detection Results on REVEAL.}
\label{tab: reveal_result}
\resizebox{\linewidth}{!}{%
\small
\begin{tabular}{lccc}\hline
Model &Prec. (\%) &Rec. (\%) &F1 (\%) \\\hline
VulDeePecker &17.7 &13.9 &15.7 \\
SySeVR &24.5 &40.1 &30.3 \\
Devign &34.6 &26.7 &29.9 \\
REVEAL &30.8 &\textbf{60.9} &41.3 \\
Transformer &41.6 &45.3 &43.4 \\
RoBERTa &44.5 &39.0 &41.6 \\ 
CodeBERT & 44.6 & 45.8 & 45.2\\
GraphCodeBERT & 47.9 & 43.9 & 45.8\\
\hline
\tool & & & \\
\mlm &45.5 &31.0 &36.9 \\
\mlmclr &38.6 &47.7 &42.6 \\
\mlmclrpn &39.4 &50.5 &44.2 \\
\tmlmclrpn &\textbf{48.3} &44.6 &\textbf{46.4} \\ 
\hline

\end{tabular}
}

\end{table}
\begin{table}[ht]\centering
\vspace{-2mm}
\caption{\small Results on CodeXGLUE for vulnerability detection}\label{tab: devign_result}
\small
\begin{tabular}{lcc}
\hline
Model &Acc (\%) \\\hline
Transformer &62.0 \\
RoBERTa &61.0 \\
CodeBERT &62.1 \\
GraphCodeBERT &63.2 \\
C-BERT & 63.6$^*$ \\ \hline
\tool & \\
\mlm & 61.8 \\
\mlmclr &\textbf{64.4} \\
\mlmclrpn &63.6 \\
\tmlmclrpn &63.8 \\
\hline
\end{tabular}
\begin{flushleft}
\scriptsize 
*We take this result from \citet{buratti2020cbert}. They did not use CodeXGLUE splits, so the test data can be different with other baselines. 
\end{flushleft}
\vspace{-2mm}
\end{table}

In our case, the best \tool variation with contrastive learning and NT-MLM objective
outperforms all the baselines, including the graph-based approaches and models pre-trained with larger datasets. This empirically proves that \tool can efficiently understand the code semantics and data dependencies from limited amount of data, helping the identification of the vulnerable patterns. We notice that hard negative samples (\ie buggy code contrasts) helps \tool improve the performance. The reason is that REVEAL contains thousands of (buggy version, fixed version) pairs for the same function. Two functions in such a pair are different by only one or a few tokens. Such real-world challenges align well with our automatically generated buggy code, and pre-training with these examples teaches the model better distinguish the buggy code from the benign ones. We provide an example in Appendix Figure~\ref{fig:reveal_example} to illustrate this. 

\noindent\textbf{CodeXGLUE.} We consider four pre-trained models: RoBERTa, CodeBERT, GraphCodeBERT and C-BERT. The first three are pre-trained on much larger datasets than ours. However, even trained with small dataset, three variations of \tool outperforms the baselines.
Unlike REVEAL, CodeXGLUE does not have those challenging pairs of functions' buggy and patched version;  thus the hard negative contrast in \tool does not help the model much.


\begin{table}[ht]\centering

\vspace{-2mm}
\caption{\small Clone detection on POJ104 and BigCloneBench}\label{tab:clone_result}
\resizebox{\linewidth}{!}{%
\small
\begin{tabular}{l|r|rrr}\hline
\multirow{2}{*}{Model} &POJ104 &\multicolumn{3}{c}{BigCloneBench} \\\cline{2-5}
&MAP@R &Prec.(\%) &Rec.(\%) &F1(\%) \\\hline
Transformer &62.11 &- &- &-  \\
MISIM-GNN &82.45 &- &- &- \\
RoBERTa &76.67 &- &- &- \\
CodeBERT$^*$ &82.67 &94.7 &93.4 &94.1\\
GraphCodeBERT$^*$ &- &94.8 &\textbf{95.2} &\textbf{95.0} \\\hline
\tool & & & & \\
\mlm &\textbf{83.32} &93.4 &93.8 &93.6\\
\mlmclr &82.44 &93.9 &93.7 &93.8\\
\mlmclrpn &82.73 &\textbf{95.1} &93.3 & 94.2\\
\tmlmclrpn &82.77 &94.2 &94.6 &94.4 \\
\hline

\end{tabular}
}
\begin{flushleft}
\scriptsize 
*The authors of both works fixed bugs in their evaluation tool and updated the results in their Github repositories. We directly take their latest results and use their latest evaluation tool for fair comparisons.
\end{flushleft}
\end{table}
\vspace{-2mm}




\subsection{Clone Detection}
\label{subsec:clone_detect}
Clone detection aims to identify the programs with similar functionality. 
It also can help detecting security vulnerabilities---given a known vulnerability, we can scan the code base with clone detector and check for similar code snippets.

\noindent\textbf{Dataset and Metrics.} We consider POJ-104~\citep{mou2016convolutional} and BigCloneBench~\citep{svajlenko2014towards} as the evaluation datasets. We again strictly follow the CodeXGLUE train/dev/test splits for experiments. Following CodeXGLUE's design, we use MAP@R as the metric for POJ-104 and precision/recall/F1 as the metric for BigCloneBench.

\noindent\textbf{POJ-104.} We consider three pre-trained models, one graph-based model~\citep{ye2020msiim} and one 12-layer Transformer model trained from scratch as baselines. Table~\ref{tab:clone_result} shows that, with hard negative contrast and NT-MLM, \tool outperforms all baselines including CodeBERT, which is pre-trained on much larger datasets.
This highlights the significance of learning the code contrasts together with syntactical information to better capture the functional similarities. Interestingly, we notice that DISCO-MLM performs the best among all variations. This indicates that our current positive heuristics might not align with all the clone patterns in this benchmark. As future work, we will propose more code transformation rules to imitate more real-world clone patterns.


\noindent\textbf{BigCloneBench.} Our best model achieves slightly better precision than the baselines indicating that our designs with contrastive learning and structure information can compensate the loss brought by less data. However, our recall is slightly worse than GraphCodeBERT, since they are pre-trained on large datasets with code graph. We conclude that enlarging our Java pre-training dataset is necessary for code clone detection and we regard this as future work.

\subsection{Medium Pre-trained Model}
\label{subsec:medium_model}
As shown in Section~\ref{sec5:expr}, \tool trained on a small dataset achieves comparable or even better performance than models pre-trained on large datasets in vulnerability and clone detection (Let's call this version  \toolS). We further explore the benefits of pre-training using larger data. We pre-train a {\sc Medium} model, \toolM,  on our extended datasets with more C-language Github repositories (13G). Note that our medium dataset is still smaller than the large dataset of the baseline models (13G vs.~20G). We evaluate \toolM on C-language tasks. The results are shown in Table~\ref{tab: medium_model}. Increasing the pre-training dataset improves the performance of downstream tasks, outperforming the best baselines' results. 

\begin{table}[h]\centering
\caption{\small Results for the best baselines, small- 
and medium- \tool. 
POJ-104 is for code clone task; 
VD-CXG and VD-RV are VD tasks for CodeXGLUE and REVEAL datasets.
}
\label{tab: medium_model}
\resizebox{\linewidth}{!}{%
\small
\begin{tabular}{lrrrr}\hline
\multirow{2}{*}{Model} & POJ-104 &VD-CXG  &VD-RV \\
 &  (MAP@R) & (Acc) & (F1) \\
 \hline
\toolS &82.8 &63.8 &46.4 \\
\toolM &\textbf{83.8} &\textbf{64.6} &\textbf{50.2} \\
Baseline$_{large}$ &82.7 & 63.6 &45.8\\
\hline
\end{tabular}
}
\end{table}





\section{Conclusion}
\label{sec:conclusion}
In this work, we present \tool, a self-supervised contrastive learning framework to both learn the general representations of source code and specific characteristics of vulnerability and code clone detections. Our evaluation reveals that \tool pre-trained with smaller dataset can still outperform the large models' performance and thus prove the effectiveness of our design.
\section*{Acknowledgements}
We would appreciate the insightful feedback and comments from the anonymous reviewers.
This work was partially done when Yangruibo Ding was an intern at IBM Research. This work is also supported in part by NSF grants CCF-2107405, CCF-1845893, IIS-2040961, and IBM.

\section*{Ethical Considerations}
The main goal of \tool is to generate functionality-aware code embeddings, producing similar representations for code clones and differentiating security bugs from the benign programs. 
Our data is collected from either the open-source projects, respecting corresponding licences' restrictions, or publicly available benchmarks. Meanwhile, throughout the paper  we make sure 
to summarize the paper's main claims. We also discussed \tool's limitation and potential future work for clone detection in Section~\ref{subsec:clone_detect}. We report our model configurations and experiment details in Appendix~\ref{appendix: model_config}.

\bibliographystyle{acl_natbib}
\bibliography{main}


\appendix
\section{Appendix}

\vspace{-1mm}
\subsection{Bug Injection Heuristics and Common Weakness Enumeration Types}
\label{appendix:bug_injection_cwe}
\vspace{-1mm}

Our automated bug injection heuristics are motivated by the real-world security bugs that are always small but hazardous. We empirically conclude the frequently happened vulnerable patterns based on the concrete CWE types. Table~\ref{tab: cwe} shows that each of our operation is relating to several CWE types. We inject all these security issues automatically and ask model to distinguish them with the benign samples.

\vspace{-2mm}
\subsection{Node Type Details}
\vspace{-1mm}
We parse the source code into ASTs and extract the node type and parent node type for each token. Table~\ref{tab: node_type} shows an example after parsing. We can see that, with the parent node type, each token can be well embedded with its local structure contexts. Considering two tokens that are distant from each other: \texttt{if} and \texttt{else}. With only node types, we just know these two tokens are keywords, but with parent node type, we can easily know that they are from the same \texttt{if-statement} and they are siblings in the AST. 

\vspace{-2mm}
\subsection{Dataset } 
\vspace{-1mm}

\noindent\textbf{Pre-training} We collect our dataset from C and Java Github repositories. Our main dataset is the combination of Java {\sc small} and C {\sc small}. From Table~\ref{tab: pretrain_data_stats}, we can see that our dataset is significantly smaller than the existing pre-trained models. For an ablation study (\S~\ref{subsec:medium_model}) with enlarged dataset, we collect a {\sc medium} dataset of C language. We have seen the improvement using such larger dataset, but even {\sc medium} dataset is still much smaller than other datasets.

\noindent\textbf{Datasets for downstream tasks} We provide dataset details of our downstream tasks in Table~\ref{tab: downstream_dataset}. Noted that for POJ-104~\citep{mou2016convolutional}, Table~\ref{tab: downstream_dataset} only shows the number of code samples, and we follow the design of CodeXGLUE that build positive and negative pairs during the minibatch generation. The amount of pairs for training is much larger than the number of samples.

\begin{table}
\centering
\caption{Comparison of pre-training dataset size between ours and other related work}
\label{tab: pretrain_data_stats}
\small

\begin{tabular}{|l|c|c|c|}
\hline
\textbf{Dataset} & \textbf{Instance Count}& \textbf{Total Size}\\
 \hline
\tool & & \\
Java {\sc small} &187 K& 992 MB \\
C {\sc small} &64 K& 865 MB \\
C {\sc medium} &860 K& 12 GB \\ \hline
CodeBERT &8.5 M & 20 GB \\
GraphCodeBERT &2.3 M & 20 GB \\
CuBERT &7.4 M &- \\
DOBF & - & 45 GB \\
PLBART &- &576 GB \\
\hline
\end{tabular}
\vspace{-3mm}

\end{table}
\begin{table*}
\vspace{-3mm}
\centering
\caption{Common Weakness Enumeration (CWE) types covered by our bug injection heuristic}\label{tab: cwe}
\small
\begin{tabular}{|l|l|}\hline
Operation &Potential CWE types \\\hline
\multirow{6}{*}{Misuse of Data Type} &CWE-190: Integer overflow \\
&CWE-191: Integer Underflow \\
&CWE-680: Integer Overflow to Buffer Overflow \\
&CWE-686: Function Call With Incorrect Argument Type \\
&CWE-704: Incorrect Type Conversion or Cast \\
&CWE-843: Access of Resource Using Incompatible Type \\\hline
\multirow{3}{*}{Misuse of Pointer} &CWE-476: NULL Pointer Dereference \\
&CWE-824: Access of Uninitialized Pointer \\
&CWE-825: Expired Pointer Dereference \\\hline
\multirow{10}{*}{Change of Conditional Statements} &CWE-120: Buffer Overflow \\
&CWE-121: Stack-based Buffer overflow \\
&CWE-122: Heap-based Buffer overflow \\
&CWE-124: Buffer Underflow \\
&CWE-125: Out-of-bounds Read \\
&CWE-126: Buffer Over-read \\
&CWE-129: Improper Validation of Array Index \\
&CWE-787: Out-of-bounds Write \\
&CWE-788: Access of Memory Location After End of Buffer \\
&CWE-823: Use of Out-of-range Pointer Offset \\\hline
\multirow{4}{*}{Misuse of Values} &CWE-369: Divide By Zero \\
&CWE-456: Missing Initialization of a Variable \\
&CWE-457: Use of Uninitialized Variable \\
&CWE-908: Use of Uninitialized Resource \\\hline
\multirow{5}{*}{Change of Function Calls} &CWE-683: Function Call With Incorrect Order of Arguments \\
&CWE-685: Function Call With Incorrect Number of Arguments \\
&CWE-686: Function Call With Incorrect Argument Type \\
&CWE-687: Function Call With Incorrectly Specified Argument Value \\
&CWE-688: Function Call With Incorrect Variable or Reference \\
\hline
\end{tabular}
\vspace{-3mm}
\end{table*}
\begin{table*}

\caption{Examples for tokens and their AST node types }
\label{tab: node_type}
\small
\centering
\vspace{-3mm}
\begin{tabular}{|l|r|r||l|r|r|}
\hline
\textbf{token} &\textbf{node type} &\textbf{parent node type} &\textbf{token} &\textbf{node type} &\textbf{parent node type} \\\hline
int &type &func\_definition &) &punctuation &parenthesized\_expr \\
foo &identifier &func\_declarator &return &keyword &return\_stmt \\
( &punctuation &param\_list &( &punctuation &parenthesized\_expr \\
int &type &parameter\_declaration &1 &number\_literal &parenthesized\_expr \\
bar &identifier &parameter\_declaration &) &punctuation &parenthesized\_expr \\
) &punctuation &parameter\_list &; &punctuation &return\_stmt \\
\{ &punctuation &compount\_stmt &else &keyword &if\_stmt \\
if &keyword &if\_stmt &return &keyword &return\_stmt \\
( &punctuation &parenthesized\_expr &( &punctuation &parenthesized\_expr \\
bar &identifier &binary\_expr &0 &number\_literal &parenthesized\_expr \\
$<$ &operator &binary\_expr &) &number\_literal &parenthesized\_expr \\
5 &number\_literal &binary\_expr &; &punctuation &return\_stmt \\
& & &\} &punctuation &compount\_stmt\\

\hline

\end{tabular}
\vspace{-3mm}
\end{table*}
\begin{table*}
\vspace{-3mm}
\centering
\caption{Details of downstream tasks datasets.}\label{tab: downstream_dataset}
\small
\begin{tabular}{|l|c|c|c|c|c|}\hline
Task &Dataset &Language &Train &Valid &Test \\\hline
\multirow{2}{*}{Vulnerability Detection} &\citet{chakraborty2021reveal} &C/C++ &15,867 &2,268 &4,535 \\
&\citet{Zhou2019DevignEV} &C/C++ &21,854 &2,732 &2,732 \\\hline
\multirow{2}{*}{Clone Detection} &\citet{mou2016convolutional} &C/C++ &32,000 &8,000 &12,000 \\
&\citet{svajlenko2014towards} &Java &901,028 &415,416 &415,416 \\\hline
\end{tabular}
\vspace{-3mm}
\end{table*}
\begin{figure*}
    \vspace{-3mm}
    \centering
    \includegraphics[width=0.95\textwidth]{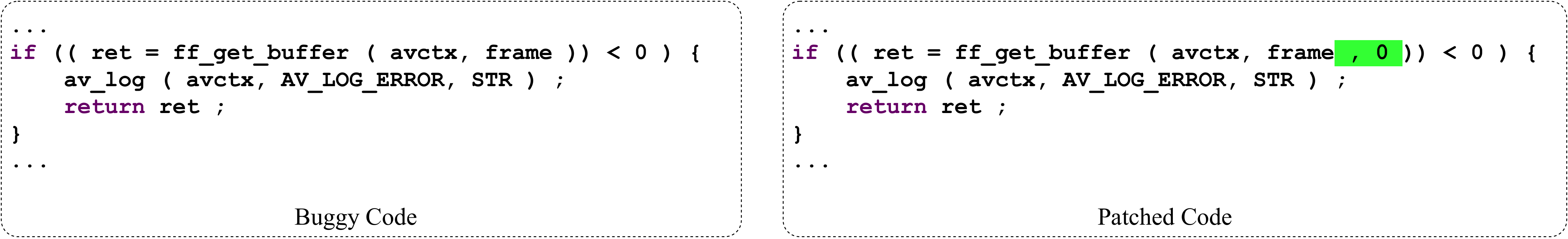}
    \caption{An example in REVEAL dataset. The patched code happens to be in the train split and the buggy code is in the test split. During inference, \tool \mlmclrpn model can correctly predict the buggy code as vulnerable, while \mlmclr predicts it as benign.}
    \label{fig:reveal_example}
    \vspace{-3mm}
\end{figure*}

\vspace{-1mm}
\subsection{Configuration }
\vspace{-1mm}
\label{appendix: model_config}
\tool is built based on a stack of 12 layers transformer encoder with 12 attention heads and 768 hidden sizes. The model is implemented with PyTorch-1.9.0 and Huggingface-transformer-4.12.3~\footnote{\url{https://github.com/huggingface/transformers/tree/c439752482759c94784e11a87dcbf08ce69dccf3}}.
Longer sequences are disproportionately expensive so we follow the original BERT design by pre-training the model with short sequence length for first 70\% steps and long sequence length for the rest 30\% steps to learn the positional embeddings. \tool is trained with Java {\sc small} and C {\sc small} for 24 hours in total with two 32GB NVIDIA Tesla V100 GPUs, using batch size of 128 with max sequence length of 256 tokens and batch size of 64 sequences with max sequence length 512 tokens.
\tool is also trained with C {\sc medium} for 3 days, using batch size of 1024 sequences with max sequence length of 256 tokens and batch size of 512 sequences and max sequence length of 512 tokens. 
We use the Adam optimizer and 1e-4 as the pre-training learning rate. For fine-tuning tasks, we use batch size of 8 and the learning of 8e-6. We train the model with train split and evaluate the model during the training using validation split. We pick the model with best validation performance for testing.

\vspace{-1mm}
\subsection{Case Study}
\vspace{-1mm}

We studied the model performance on REVEAL dataset for vulnerability detection. Figure~\ref{fig:reveal_example} shows two samples inside REVEAL. We can recognize that they are from the same program. We further checked the details of these two example and we found the code on the left is a buggy version, and it is fixed by adding an argument of value 0 to the function call. This real-world situation actually matches our "Change of Function Calls" (\S~\ref{subsec:bug_injection}) bug injection operation. In the REVEAL dataset, the patched code is in the training corpus while the buggy one is in the test split. Interestingly, during the inference, \tool \mlmclrpn can correctly predict the buggyiess while \mlmclr fails. This empirically prove our bug injected samples can help the model identify small but siginicant real-world vulnerabilities.

\end{document}